\documentclass[journal]{IEEEtran}

\normalsize

\ifCLASSINFOpdf
\else
\fi

\usepackage{verbatim} 
\usepackage[dvips]{graphicx} 
\usepackage[cmex10]{amsmath}
\usepackage{amssymb,amsfonts}
\usepackage{mathtools}  
\usepackage{framed}
\usepackage{lipsum}
\usepackage{mdframed}
\usepackage{adjustbox}

\usepackage{cases}
\usepackage{theorem} 
\def\myproof{\noindent{{\textbf{Proof:}}}} 

\ifdefined\myproof
\else
\def\myproof{\proof}

\fi

\newtheorem{remark}{Remark}

\newtheorem{proof}{Proof} 

\usepackage{cite} 

\usepackage[tight,footnotesize]{subfigure}

\usepackage[linesnumbered,ruled]{algorithm2e}
\usepackage{multirow}

\begin{document}

%
\title{ Effect of Antenna Deployment on Achievable Rate in  Cooperative Magnetic Induction Communication}
%
%
%
%

\author{Honglei~Ma,
        Erwu~Liu, ~\IEEEmembership{Senior Member,~IEEE,}
        Rui~Wang, ~\IEEEmembership{Member,~IEEE,}
        and Xinyu~Qu
 \thanks{This work was published in IEEE Communications Letters, with the DOI: 10.1109/LCOMM.2019.2929790.}
\thanks{
This work is supported by National Science Foundation China under Grants (61571330 and 61771345).  \emph{(Corresponding author: Erwu Liu.)}  }
\thanks{
 Honglei Ma, Erwu Liu, Rui Wang and Xinyu Qu are with the School of
Electronics and Information Engineering, Tongji University, Shanghai, China (e-mail: holyma@yeah.net; erwu.liu@ieee.org; ruiwang@tongji.edu.cn; 1410459qxy@tongji.edu.cn).}
}

%
%

\markboth{}%
{Submitted paper}
%



\maketitle

\begin{abstract}
Magnetic Induction (MI) communication can be applied in some through-the-earth scenarios such as mines and underground rivers. To increase the transmission rate of MI communications, we propose a cooperative MI (CMI) scheme with an amplify-and-forward (AF) relay. Different from existing studies, we mainly focus on the relay with arbitrary antenna position and orientation (antenna deployment, AD).  We derive the closed-form expression of CMI achievable data rate gain (CMG) for the relay and the closed-form expression of CMI channel bandwidth. Simulations reveal that a relay with appropriate AD could yield a significant increase in the achievable rate for MI systems.
\end{abstract}


%
\IEEEpeerreviewmaketitle

\section{Introduction}\label{sect_intro}
%
%
%
%

\IEEEPARstart{T}{he} magnetic induction (MI) communication turns out to be a reliable communication technology for the through-the-earth (TTE) environments\cite{akyildiz2009signal} where  electromagnetic (EM) wave extremely attenuates through underground media which is hundreds of meters thick. Using 8m transmit antenna, Lockheed Martin and Chinese researchers successfully record a longest propagation range of 472.5m and 310m, respectively\cite{Zhang2014Cooperative}.  Although researchers study optimization problems for the MI capacity improvement\cite{kisseleff2013channel}\cite{Sun2013Increasing},  the transmission rate is extremely low for the 1--100 KHz operating frequency.

The idea of MI multiple-input multiple-output (MIMO) has been proposed in \cite{Li2015Capacity}. However, the crosstalk among MIMO coils is serious. The MIMO antennas are difficult to be deployed to avoid antennas crosstalk\cite{Kim2016Field} in the narrow underground passages due to the size of antennas.

The cooperative relay techniques in the near field magnetic induction communications (NFMIC) are investigated recently. The MI waveguide technique is a passive relay method and can provide range extension and data rate improvement\cite{Sun2010Magnetic}. Based on MI waveguide methods, researchers find that the configuration of the passive relaying coils which are placed on the same vertical axis maximizes the receive power\cite{Masihpour2010Cooperative}.

Unfortunately, the literature\cite{Kisseleff2015On} shows that the passive relay provides only a limited performance improvement and investigates the potential of an active relay device. Based on a single waveguide mode, the authors of \cite{Kisseleff2015On} assume a single relaying waveguide model and analyze the properties of different active relaying protocols such as AF and decode-and-forward (DF). Also, they optimize the power allocation and signal frequency for the CMI scheme.  Practically, the leased relay nodes may be located at arbitrary positions. However, the CMI performance\emph{ w.r.t. }coils position and orientation (called \emph{antenna deployment}, AD) is not analyzed.

 To the best of our knowledge, this paper is the first attempt to investigate the connection between the AD and the CMI data rate. Our proposed CMI network model with an AF relay is different from the most existing CMI models \cite{Masihpour2010Cooperative}\cite{Kisseleff2015On} which are similar to MI waveguide networks. Firstly, any two related nodes can communicate with each other directly. Secondly, this letter focuses on the TTE scenarios with large distance, low frequency and remarkable eddy-current loss. Thirdly, the relay node is with arbitrary AD in a CMI network where the antennas (coils) of source and destination are not assumed to be parallel.

Therefore, our contributions include two parts: 1) Using methods of the approximation and variable substitution, we derive the closed-form expression of CMI channel bandwidth which is mathematically intractable. We find that the transmit power and ADs have a significant effect on CMI channel bandwidth; 2) By using magnetostatic field theorems and dimensional rotation group  techniques, we derive the closed-form expression of CMG \emph{ w.r.t.} AD which can guide the relay selection. From the CMG expression and simulation results, we analyze the potential optimal AD in most TTE scenarios.

       \begin{figure}[b]
        \centering
        \includegraphics[width=2.9in]{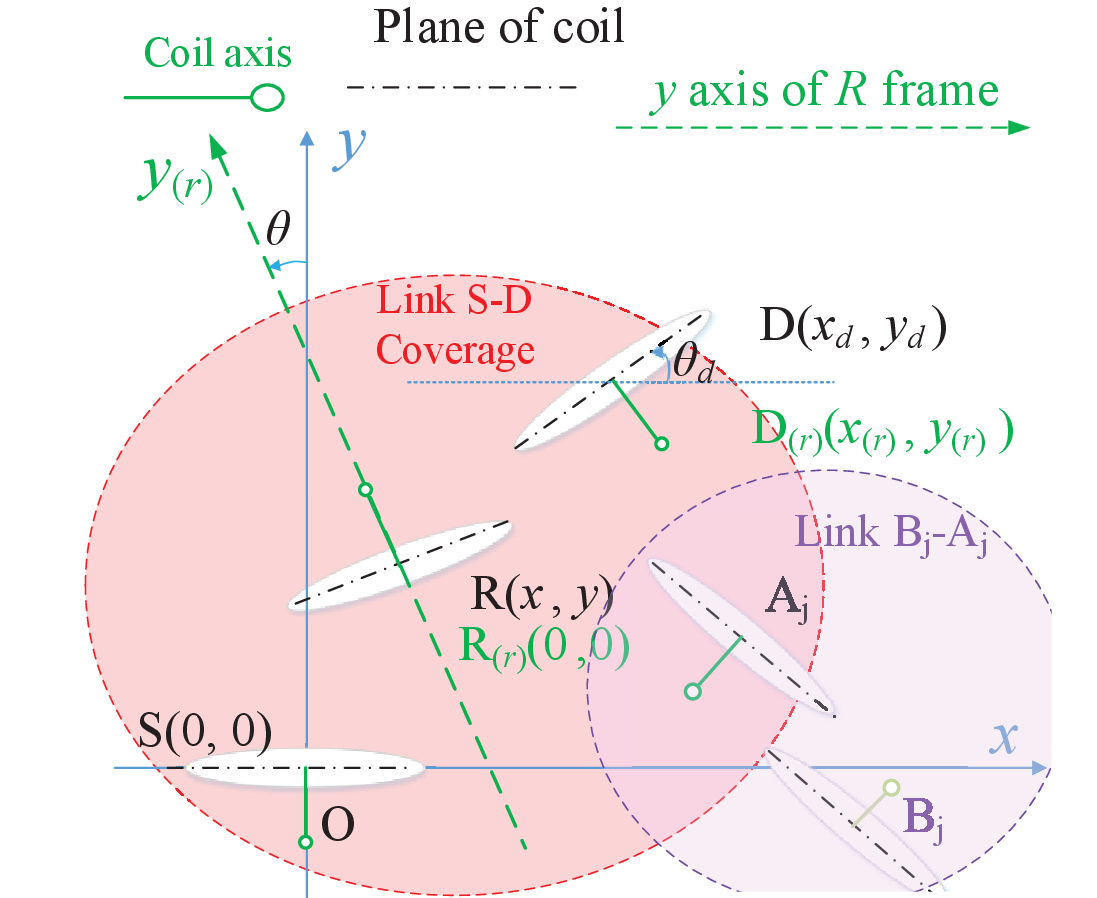}\\
        \caption{A cooperative MI network with a single relay. The subscript((r)) denotes frame R.
 }\label{miAngle}
    \end{figure}

   \begin{figure}[ht]
        \centering
        \includegraphics[width=3.1in]{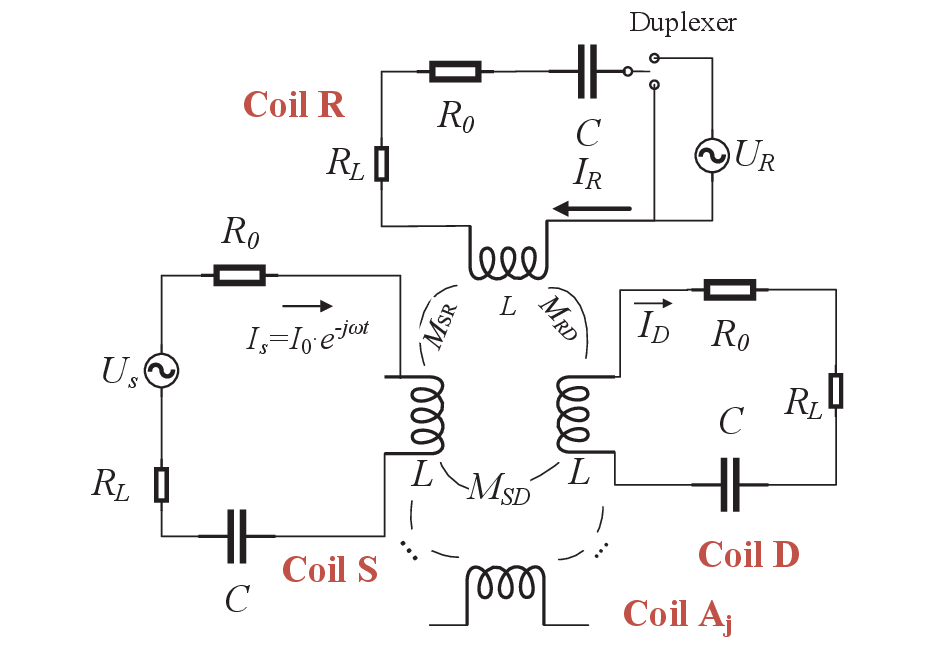}\\
        \caption{Equivalent circuit model for the cooperative MI network.
 }\label{miCircle}
    \end{figure}

\section{System Model}\label{sect model}
In this section, we introduce a CMI network for TTE scenarios and derive the expressions of mutual inductance \emph{ w.r.t.} the AD.
\subsection{CMI Systems}\label{subsect_model_sys}
    We propose a CMI communication system model comprising an MI source node S with AD $O= (0, 0, 0)$, an MI relay node R with arbitrary AD $\mathbf{v}= (x, y, \theta)$ and an MI destination node D with specific AD $\mathbf{v}_d= (x_d, y_d, \theta_d)$ as depicted in Fig.\ref{miAngle}. Here $(x, y)$ represents the position and $\theta$ is the antenna (coil) orientation which is the angle between the x axis and the radial direction of the coil. For the existence of $B_j$--$A_j$ link, the coil $A_j$ within the signal coverage of S-D link might become an unexpected relay, \emph{i.e.,} the relay whose coil might form unexpected additional resonance circuits with other working node coils. The node $A_j$ is with the arbitrary AD $\mathbf{v}_{aj}$= ($x_{aj}$, $y_{aj}$, $\theta_{aj}$).
    Similar to \cite{Kisseleff2015On}, each coil has the radius $a$, the number of turns $n$, the inductivity $L$ and the resistance $R_0$. The capacitor with capacitance $C$  is tuned to make the circuit resonant at frequency $f_0$ =$1/2\pi\sqrt{LC}$, and $R_L$ is load resistor, see Fig.\ref{miCircle}. For the TTE scenarios, we assume that $f_0$ is small sufficiently for the eddy-current loss reduction, and the minimal distance $d_0$ between two nodes is sufficiently large, \emph{i.e.}, $d_0 > 4a$.

    Hereafter, a working node set within the signal coverage of S-D link is designated as $\mathcal{N}=\{S, R, D, A_j\}$. The notation $M_{kl}$ represents the mutual inductance between node $k$ and node $l$. The notation $|H_{kl}|^2$ denotes the channel power gain between two nodes $k$ and $l$.

\subsection{Mutual Inductance and Antenna Deployments}\label{subsect_model_optm}
For the MI ADs in the multi-user network, it is most important to express the mutual inductance under a common Cartesian frame. Most existing studies focus on the relative position between two points, \emph{e.g.}, based on \cite{balanis2005antenna},  the field at an arbitrary point is approximately expressed under the spherical frame with the magnetic field components $\textbf{H}_{(r)}\left(r,\theta_{(r)}\right) \! = \! \left[\frac{n a^2 I_0\cos\theta_{(r)}}{2 r^3}\hat{e}_r + \frac{n a^2 I_0\sin\theta_{(r)}}{4 r^3}\hat{e}_\theta\right] e^{-r/\delta}$
where  $I_0$ is transmitting current, $\delta$ is the skin depth. Thus, the magnetic field at D under the Cartesian frame R is deduced as
\begin{equation}\label{eqn_Hs_loopr}
\begin{aligned}
\textbf{H}_{(r)}\left(x_{(r)},y_{(r)}\right) = \frac{n I_0 a^2}{4 e^{\frac{r}{\delta}}} \left( \frac{2y_{(r)}^2 - x_{(r)}^2}{x_{(r)}^2+y_{(r)}^2}, \frac{3 x_{(r)}y_{(r)}^2 }{x_{(r)}^2+y_{(r)}^2} \right)^T,
\end{aligned}
\end{equation}
where the superscript $(\cdot)^T$ denotes transpose. We rotate and translate from frame R to frame S  and get
 \vspace{-0.5em}
\begin{subequations}\label{eqn_Hs_loop}
\begin{align}
   \begin{pmatrix}x_{(r)}\\y_{(r)}\end{pmatrix} &= \begin{bmatrix}\cos\theta & \sin\theta\\ -\sin\theta&\cos\theta \end{bmatrix} \begin{pmatrix}x_{d} - x\\y_{d}-y \end{pmatrix}, \label{eqn_Hs_loop:a}
   \\ \textbf{H}(x_d, y_d) &= \begin{bmatrix}\cos\theta & \sin\theta\\ -\sin\theta&\cos\theta \end{bmatrix}^{-1} \textbf{H}_{(r)}. \label{eqn_Hs_loop:b}
\end{align}
\end{subequations}
 By substituting~\eqref{eqn_Hs_loopr} and~\eqref{eqn_Hs_loop:a} into~\eqref{eqn_Hs_loop:b}, we get the expression of $\textbf{H}(\mathbf{v}, \mathbf{v}_d)$. Thus, the mutual inductance between arbitrarily placed coils R and D \emph{w.r.t.} their ADs $\mathbf{v}$ and $\mathbf{v}_d$ is $\partial\Phi/\partial I_0$, \emph{i.e.},

\begin{equation}\label{eqn_mrd}
\begin{aligned}
  & M_{RD}(\mathbf{v}, \mathbf{v}_d)= \frac{\partial\left(n\mu\pi I_0 a^2 e^{-\frac{r}{\delta}}\left|\textbf{H}(\mathbf{v},\mathbf{v}_d)\cdot\textbf{n}_d(\mathbf{v}_d) \right| \right)}{\partial I_0}=  \\
  &  \frac{\splitfrac{ \left(\Delta x^2+\Delta y^2\right)^2  \cos(\theta_d-\theta) -6\Delta x\Delta y\sin(\theta_d+\theta)}
   {- 3  \left( \Delta x^2 -\Delta y^2\right)^2 \cos(\theta_d+\theta)}}
   {2\kappa^{-1} \cdot e^{\sqrt{\Delta x^2+\Delta y^2}/\delta}\sqrt{\left(\Delta x^2+\Delta y^2\right)^5}},
  \end{aligned}
\end{equation}
where $\Delta x=(x_d-x)$, $\Delta y=(y_d-y)$, \textbf{n}$_d(\mathbf{v}_d)$  is the normal vector of the receiving coil, $\kappa$ = $\mu$$\pi$\emph{n}$^2a^4 /4$, $\mu$ denotes the medium permeability and $\Phi$ denotes magnetic flux. Therefore, $M_{SR}$ and $M_{SD}$ are  respectively  deduced as
\begin{equation}\label{eqn_msd}
\begin{aligned}
   & M_{SR}(\mathbf{v}) = M_{RD}(\mathbf{0}, \mathbf{v})=\frac{\left( 2y^2-x^2\right) \cos \theta  - 3x y\sin \theta }
   {2\kappa^{-1}e^{\frac{\sqrt{ x^2+ y^2}}{\delta}} \sqrt{\left(x^2+y^2\right)^5}},  \\
  & M_{SD} =  M_{SR}(\mathbf{v}_d)=  M_{SR}(x_d, y_d, \theta_d).
\end{aligned}
\end{equation}
Also, we obtain $M_{RA_j}$$=$$M_{RD}(\mathbf{v}, \mathbf{v}_{aj})$ and $M_{SA_j}$$=$$\!M_{SR}(\mathbf{v}_{aj})$.  For an MI relay network with specific source and destination,  $x_d$, $y_d$ and $\theta_d$ are treated as constants. As a result, $M_{sd}$ is treated as a constant.

\section{ Achievable Rate and Antenna Deployment  }\label{sect capa}
In this section,  the signal-noise ratio (SNR) is analyzed. The mathematically intractable expression of CMI bandwidth is deduced, followed by the derivation of  CMG in comparison with the S-D link. The CMG can be used for relay selections.
\subsection{Channel Coefficients and SNR}\label{subsect_model_gain}
 According to the principle of cooperative communication, the signal \emph{s} generated by a power supply with voltage $U_S$ is transmitted from S to D in two time slots. During the first time slot, the source broadcasts \emph{s} to the destination and relay. The circuit impedance of a transceiver is
 $Z_{tr}$$=$$Z_{lc}$$+$$R_0$$+$$R_L$ where
\begin{equation}\label{eqn_zlc}
\begin{aligned}
     Z_{lc}=  j2\pi fL+\frac{1}{j2\pi fC} = j(\frac{f}{2\pi f_0^2 C}- \frac{1}{2\pi f C}).
\end{aligned}
\end{equation}
Similarly to the methods given in \cite{Kisseleff2015Beamforming}, the current $I^{(1)}_k$ for each receiver coil $k\in \mathcal{N} \verb|\| \{S\}$  is determined by Kirchhoff's voltage law equations
  \begin{equation}\label{eqn_Kf1}
\begin{aligned}
   &I_{S}^{(1)}Z_{tr}+ N_{S} + \sum\limits_{l\in \mathcal{N} \verb|\| \{S\}} I_{l}^{(1)}\cdot j2\pi f M_{Sl} = U_S,  \\
   & I_{S}^{(1)} j2\pi f M_{Sk} + \sum\limits_{l\in \mathcal{N} \verb|\| \{S,k\}} I_{l}^{(1)} j2\pi f M_{kl} = -U_{Sk},   \\
\end{aligned}
\end{equation}
where $U_{Sk}\!=\!I_{k}^{(1)}Z_{tr}+ N_{k}$ and $N_{k}$ denotes the additive white Gaussian noise (AWGN) whose average power spectral density (PSD) at the receiver coil $k$ is $P_{Nf}$. The noise model is similar to that of \cite{Sun2013Increasing}. Considering the TTE scenarios with the sufficiently large distance and low frequency, the impendence caused by the mutual inductance between two nodes is much smaller than the total antenna impedance so that  $Z_{tr}\!\gg\! j2\pi f M_{Sk}$, $Z_{tr}\!\gg\!j2\pi f M_{kl}$,  $Z_{tr}\!\gg j2\pi\!f M_{Sl}$ and $U_S \gg N_{k}$. We solve~\eqref{eqn_Kf1} and obtain the current of each receiver antenna $I^{(1)}_k \simeq \frac{j2\pi f M_{Sk} Z_{tr}^2 U_S }{Z_{tr}^4}$ where $k\in \mathcal{N} \verb|\| \{S\}$. Subsequently, by substituting $I_k^{(1)}$, the channel power gains can be deduced as
 \begin{equation}\label{eqn_hsr}
\begin{aligned}
   &|H_{SR}(\mathbf{v},f)|^2 \!=\frac{|I_{k=R}^{(1)}|^2 R_L}{P_{Sf}(f)}\! \simeq \! \frac{(2\pi f)^2 R_L}{ \left|Z_{tr}(f)\right|^3}  \cdot  M_{SR}^2(\mathbf{v}),    \\
   &|H_{SD}(f)|^2  = \frac{|I_{k=D}^{(1)}|^2 R_L}{P_{Sf}(f)} \simeq\frac{(2\pi f)^2 R_L}{ \left|Z_{tr}(f)\right|^3} \cdot M_{SD}^2,
\end{aligned}
\end{equation}
  where $P_{Sf}(f)$$=$$\frac{\!U_S^2}{|Z_{tr}(f)|}$ is the transmit PSD function at the source and $(|I_{k}^{(1)}|^2 R_L)$ represents the receive PSD at node $k$. The equations~\eqref{eqn_hsr} indicate that the channel power gain between two nodes is independent with other nodes in the TTE scenarios.  {The signal-noise ratio (SNR) at the destination is $\Upsilon_1\!=\!\frac{P_{Sf}(f)}{P_{Nf}}|H_{SD}(f)|^2$\cite{Kisseleff2015On}, \emph{i.e.,}
  \begin{equation}\label{eqn_snr1}
\begin{aligned}
  \Upsilon_1(f)  = \frac{P_{Sf}(f)}{P_{Nf}}\frac{(2\pi f)^2 R_L}{ \left|Z_{tr}(f)\right|^3}  M^2_{SD}.
\end{aligned}
\end{equation}
}

 During the second time slot, the relay R becomes a transmitter and the source is treated as a potential unexpected relay. The relay R forwards signals with voltage $U_R$. Thus, by solving the Kirchhoff's voltage law equations similarly to~\eqref{eqn_Kf1}, we obtain the current $I^{(2)}_k \simeq \frac{j2\pi f M_{Rk} Z_{tr}^2 U_R }{Z_{tr}^4}$ at coil $k$ and
 \begin{equation}\label{eqn_hrd}
\begin{aligned}
 |H_{RD}(\mathbf{v},f)|^2 \!=\frac{|I_{k=D}^{(2)}|^2 R_L}{P_{Rf}(f)}\! \simeq \! \frac{(2\pi f)^2 R_L}{ \left|Z_{tr}(f)\right|^3}  \cdot  M_{RD}^2(\mathbf{v}),
\end{aligned}
\end{equation}
where $P_{Rf}(f)$$=$$\frac{\!U_R^2}{|Z_{tr}(f)|}$ is the transmit PSD function at the relay. The SNR at the destination in the second time slot is\cite{Kisseleff2015On}
 \begin{equation}\label{eqn_snr2}
\begin{aligned}
\Upsilon_2 = \frac{P_{Sf}(f) \beta^2 |H_{SR}|^2|H_{RR}|^2|H_{RD}|^2}{\beta^2|H_{RR}|^2|H_{RD}|^2P_{Nf} + P_{Nf} },
\end{aligned}
\end{equation}
where $\beta$ is the amplification coefficient and $|H_{RR}|^2 \!=\! |\frac{Z_{tr}}{Z_{tr}^{2}+(2 \pi f)^{2}\left(M_{S R}^{2}+M_{R D}^{2}\right)}| R_{L}\!\simeq\!\frac{R_L}{ |Z_{tr} |}$ is an additional channel gain.  Based on \cite{Kisseleff2015On} and \cite{Hammerstrom2006On}, the amplification coefficient can be optimized by
 \begin{equation}\label{eqn_beta}
\begin{aligned}
\beta(f) = \sqrt{P_{Rf}(f)/(P_{Sf}(f)|H_{SR}|^2 + P_{Nf})|H_{RR}|^2}.
\end{aligned}
\end{equation}
By using the optimal approach of maximum ratio combining (MRC), the SNR of the CMI network is $\Upsilon_{\text{mrc}}$ =  $\Upsilon_1 +\Upsilon_2$\cite{Kisseleff2015On}. Let $|H_0(f)| = \frac{(2\pi f)^2 R_L}{ \left|Z_{tr}(f)\right|^3} =  \dfrac{(2\pi f)^2 R_L}{ |(j2\pi fL+\frac{1}{j2\pi fC})+R_L + R_0|^3} $, after substituting~\eqref{eqn_hsr}--~\eqref{eqn_beta} into $\Upsilon_{\text{mrc}}$ yields:
 \begin{equation}\label{eqn_snr22}
\begin{aligned}
&\Upsilon_{\text{mrc}}(f,\mathbf{v})\!= \frac{P_{Sf}(f)}{P_{Nf}} H_0(f)  M^2_{SD}  +   \\
     & \frac{\frac{P_{Rf}(f)}{P_{Nf}} |H_0(f)|^2 M_{RD}^2(\mathbf{v})  \cdot \frac{P_{Sf}(f)}{P_{Nf}} M_{SR}^2(\mathbf{v}) }{\frac{P_{Rf}(f)}{P_{Nf}}|H_{0}(f)|M_{RD}^2(\mathbf{v})  +  \frac{P_{Sf}(f)}{P_{Nf}}|H_{0}(f)|M_{SR}^2(\mathbf{v})  + 1}.
\end{aligned}
\end{equation}

\vspace{-1.5em}
\subsection{CMI Channel Bandwidth}\label{subsect capa_BW}
Shannon's theorem and power reflections $(j2\pi fL+\frac{1}{j2\pi fC})$ of MI circuits indicate that the bandwidth influences the achievable rate performance of MI communication. Moreover,  from~\eqref{eqn_snr22}, we find that the relay AD affects the bandwidth significantly. However, the expression of CMI channel bandwidth is not found in existing literatures. In general, it is extremely difficult to obtain the CMI bandwidth even if the mathematical software is used. Alternatively, for the TTE scenarios, we can apply the approximation and variable substitution to deduce the bandwidth. Given a constant $m\leq2$, to get $-10\lg(\frac{1}{m})$-dB bandwidth,  we observe that  $Z_{lc}(f)\!= \!j(\frac{f}{2\pi f_0^2 C}\!-\! \frac{1}{2\pi f C})$  has a dominant effect on $\partial\Upsilon/\partial f$  where $\Upsilon \in \{\Upsilon_1, \Upsilon_{\text{mrc}}\}$ since $\frac{1}{2\pi f_0^2 C} \! \gg \! 2\pi  M_{SD}$ at the low frequency.  We can treat  $\Upsilon$ as the function of $|Z_{tr}^3|$. By solving the SNR equation $\Upsilon(|Z_{tr}^3|)\! = \!\frac{1}{m}\Upsilon(f_0)$ \emph{w.r.t.} $|Z_{tr}^3|$,  we get the positive solution $|Z_{tr}^3| = \mathcal{Z}_{c}$. Since $|Z_{tr}^3|$ is the function of frequency $f$, we solve the equation $|Z_{tr}(f)^3| \! = \!\mathcal{Z}_{c}$ and obtain two positive solutions which are $f_1$ and $f_2$. The bandwidth is $B_{w} = |f_1-f_2|$,\emph{ i.e.,}
 \begin{equation}\label{eqn_bw}
\begin{aligned}
\varpi (\mathcal{Z}_c) &= 2 \pi ^2 C^2 f_0^4 \mathcal{Z}_c^{-\frac{2}{3}}-8 \pi ^2 C^2 f_0^4 R_0^2+f_0^2, \\
  \varrho(\mathcal{Z}_c) &=  \sqrt{\left[ f_0^2-2 \pi ^2 C^2 f_0^4 \left(4 R_0^2-\mathcal{Z}_c^{-\frac{2}{3}}\right)\right]^2- f_0^4}, \\
    B_{w}(\mathcal{Z}_c) &= \sqrt{\varpi(\mathcal{Z}_c)+\varrho(\mathcal{Z}_c) }-\sqrt{\varpi(\mathcal{Z}_c)-\varrho(\mathcal{Z}_c). }
\end{aligned}
\end{equation}
The expression of the S-D link bandwidth is $B_1 = B_{w}([\frac{1}{m}(R_0+R_L)]^3)$ derived from $|\Upsilon_1(f)|\!=\!\frac{1}{m}|\Upsilon_1(f_0)|$. By solving the equation $\Upsilon_{\text{mrc}}(f)\!=\!\frac{1}{m}\Upsilon_{\text{mrc}}(f_0)$, we deduce the CMI channel bandwidth as $B_{AF}\!= \!B_{w}(\mathcal{Z}_{c,AF})$ where

 \begin{equation}\label{eqn_baf}
\begin{aligned}
  \mathcal{Z}_{c,AF} &= \frac{\splitfrac{\splitfrac{\bigl\{\hbar^2 A_0^2 \left(M_{RD}^2+M_{SR}^2\right)^2+2 \hbar A_0 \cdot }{\bigl[M_{SD}^2 \left(M_{SR}^2 + M_{RD}^2\right)
 +2 M_{SR}^2 M_{RD}^2 \bigl]}}{+M_{SD}^4 \bigl\}^\frac{1}{2}+ \hbar A_0 \left(M_{RD}^2+M_{SR}^2\right)-M_{SD}^2}}{2 \hbar \left[M_{SD}^2 \left(M_{SR}^2+M_{RD}^2\right)+M_{SR}^2 M_{RD}^2\right]},
\end{aligned}
\end{equation}
$\hbar$=$(2\pi f_0)^2 R_L\frac{P_{Sf}}{P_{Nf}}$ and $A_0$=$\frac{\Upsilon_\text{mrc}(f_0)}{m\hbar}$. From $B_1$=$B_{w}([\frac{1}{m}(R_0+R_L)]^3)$ and~\eqref{eqn_baf}, we find that the AD and transmitting power have little effect on the bandwidth of the one-hop communication and have significant effects on CMI channel bandwidth. Hence, $B_1$ can be treated as a constant.
\vspace*{-0.8em}
\subsection{CMI Achievable Rate Gain}\label{subsect capa_CMG}
  For a fair comparison between the two network channels, the transmit power of each CMI transmitter and each direct MI (DMI) transmitter are set to $P_S$ and $2P_S$, respectively. For the TTE scenarios with sufficiently long communication distance and low signal frequency,  the achievable rates of the CMI network and DMI network are given by
\begin{subequations}\label{eqn_caf}
\begin{align}
   C_{AF} &(\mathbf{v})= \frac{1}{2}\int_{f_0-\frac{1}{2}B_{AF}(\mathbf{v})}^{f_0+\frac{1}{2}B_{AF}(\mathbf{v})} \log_2(1 + \Upsilon_{\rm mrc}(f))df \notag\\
       \simeq &\tfrac{1}{2}B_{AF}(\mathbf{v}) \log_2(1 + \Upsilon_{\rm mrc}(f_0+\tfrac{1}{4}B_{AF}(\mathbf{v}),\mathbf{v})), \label{eqn_caf:af}\\
    C_{SD}&= \int_{f_0-\frac{1}{2}B_{1}}^{f_0+\frac{1}{2}B_{1}} \log_2(1 + 2\Upsilon_{1}(f))df \notag \\
            &\simeq B_{1} \log_2(1 + 2\Upsilon_{1}(f_0+\tfrac{1}{4}B_{1})), \label{eqn_caf:sd}
\end{align}
\end{subequations}
respectively.   According to\cite{Sun2013Increasing}, the in-band  spectral efficiency slightly decreases as $f$ deviates from $f_0$, and the integrals in~\eqref{eqn_caf} can be avoided by selecting a single frequency in calculating the spectral efficiency at this frequency close to the average spectral efficiency inside the bandwidth.

We define the CMG as $\mathcal{G}(\mathbf{v})$ = $C_{AF}/C_{SD}$ to assess the performance improvement of CMI system. After substituting~\eqref{eqn_hsr}-\eqref{eqn_hrd},~\eqref{eqn_snr22} and~\eqref{eqn_caf} into $\mathcal{G}(\mathbf{v})$, the closed-form expression of the CMG can be deduced as
 \begin{equation}\label{eqn_gc}
\begin{aligned}
   &\mathcal{G}(\mathbf{v})=C_{AF}/C_{SD}\simeq \zeta_B(\mathbf{v}) \cdot \\
    &\frac{ \log_{2}\left( 1 \! + \frac{1+1/m}{2}\!\left[\gamma_0  M^2_{SD} \! +\frac{M_{SR}^2(\mathbf{v})M_{RD}^2(\mathbf{v}) \gamma_0)}{ M_{SR}^2(\mathbf{v})\!+ M_{RD}^2(\mathbf{v})+ 1/\gamma_0}\right]\right) }{{2\log_{2}(1 \! + \frac{1+1/m}{2} 2\gamma_0  M^2_{SD}})},
\end{aligned}
\end{equation}
where $m\!=\!2$ (as widely used 3-dB bandwidth), $\gamma_0$ = $\frac{(2\pi f_0)^2R_L}{(R_0 + R_L)^3}\cdot \frac{P_S}{P_N} $, $P_N$ denotes the noise power and $\zeta_B(\mathbf{v}) = \frac{B_{AF}(\mathbf{v})}{B_1}$ which we call \emph{CMI bandwidth gain}. The CMG above (below) 1 indicates the data rate increasing (decreasing). The non-convex function $\mathcal{G}(\mathbf{v})$ could clearly give the limited information of the effect of AD on achievable rate as follows.

 \begin{remark}\label{prop_optv}
Considering  sufficiently large or small transmit power $P_S$ and according to~\eqref{eqn_bw} and~\eqref{eqn_baf}, $\zeta_B$ tends to a constant since  $\lim\limits_{P_S\rightarrow \infty}\mathcal{Z}_{c,AF}$$=$$\frac{A_0}{2}$ and $\lim\limits_{P_S\rightarrow 0}\mathcal{Z}^{-\frac{2}{3}}_{c,AF} $$=$$0$.
 For each AD $\mathbf{v}_i$ where $i$ is a positive integer, there exists a constant $\wp_i$ such that $\mathbf{v}_i$$\in$$\mathcal{V}_i$$=$$\{\mathbf{v} | {M_{SR}^2(\mathbf{v})}\!+\!M_{RD}^2(\mathbf{v})$$=$$\wp_i \}$.  Based on the fundamental inequalities,  $\Upsilon_{2}$$=$$\frac{M_{SR}^2(\mathbf{v}_i)M_{RD}^2(\mathbf{v}_i) \gamma_0}{ M_{SR}^2(\mathbf{v}_i)+ M_{RD}^2(\mathbf{v}_i)+ 1/\gamma_0} \leq \frac{\frac{1}{4}\gamma_0\wp^2_i}{\wp_i + 1/\gamma_0}$  where the equation holds when $M^2_{SR}(\mathbf{v}_i^*)$$=$$ M^2_{RD}(\mathbf{v}_i^*)$, i.e., $\mathcal{G}(\mathbf{v}_i^*)$ is optimal for all $i$.
 \end{remark}

Especially, when the antenna orientation of D $\theta_d$ tends to 0, the optimal relay position tends to be closed to the midpoint of S-D.

 For practical use, we give the definition of the \emph{valid relay area} (RA) where any relay with suitable coil orientation can yield an increase in achievable rate. In addition,
 the receiver SNR at the destination is greater than a certain threshold $\Upsilon_{th}$. For instance, to guarantee correct 2-PSK demodulation, the SNR at the destination is above $(\text{erfc}^{-1}(2\cdot \text{BER}))^2\!=\!(\text{erfc}^{-1}(2\times 10^{-3}))^2$ = 4.7748 where $\text{erfc}^{-1}$ is the inverse of Gaussian error function\cite{Sun2010Magnetic}\cite{Kisseleff2014}.

Based on the CMG and RA, we propose the relay selection procedures as follows: 1) When the network is initiated, the source calculates the RA; 2) The source selects the relay with maximal CMG  from all the idle nodes in the RA when necessary. Especially, for the link where the antennas of the source and destination are nearly parallel ($|\theta_d|< 10^\circ$), the source could prefer the idle node close to the midpoint between the source and destination; 3) The selected node adjusts the antenna orientation and works as a relay.

\section{Simulation}\label{sect sim}

   \begin{figure}[t]
        \centering
        \includegraphics[width=3.2in]{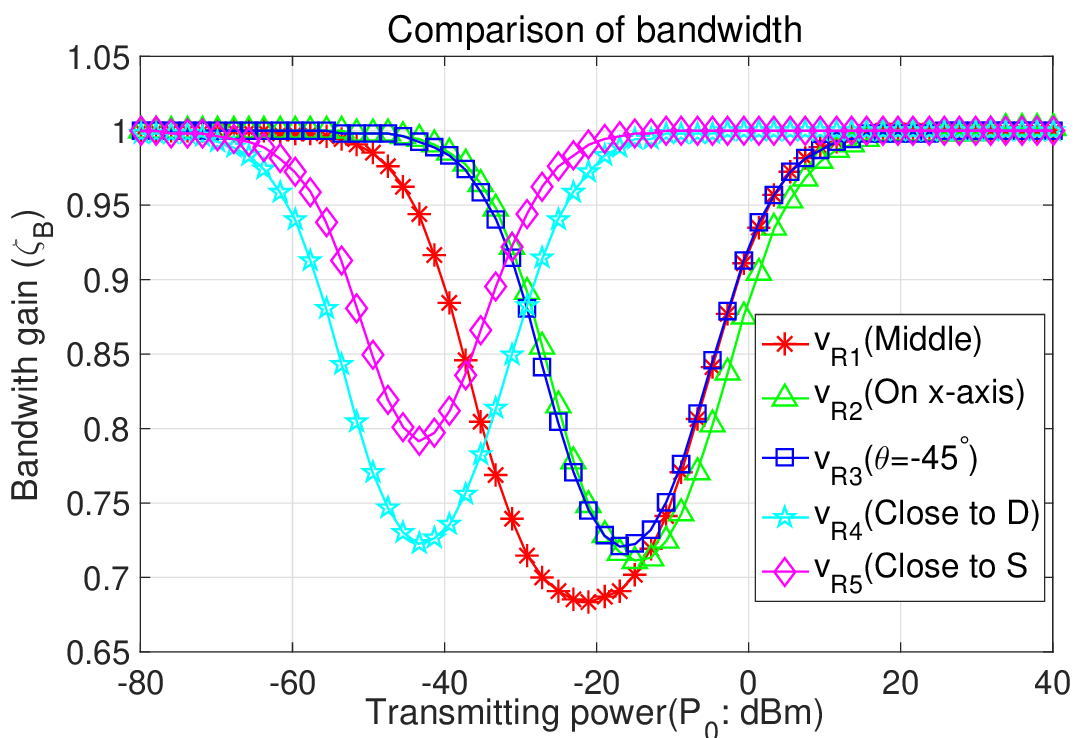}\\
        \caption{The effects of AD and transmitting power on CMI 3-dB bandwidth for the 5 relay ADs, \emph{i.e.,} $\mathbf{v}_{R1}= (25, 43.3, 0)$, $\mathbf{v}_{R2}= (50, 0, 0)$, $\mathbf{v}_{R3}= (0, 86.6, -45^\circ)$,  $\mathbf{v}_{R4}= (40, 76.6, 0)$ and $\mathbf{v}_{R5}= (10, 10, 0)$.
 }\label{fig_bandwithP00}
    \end{figure}

   \begin{figure}[t]
        \centering
        \includegraphics[width=3.4in]{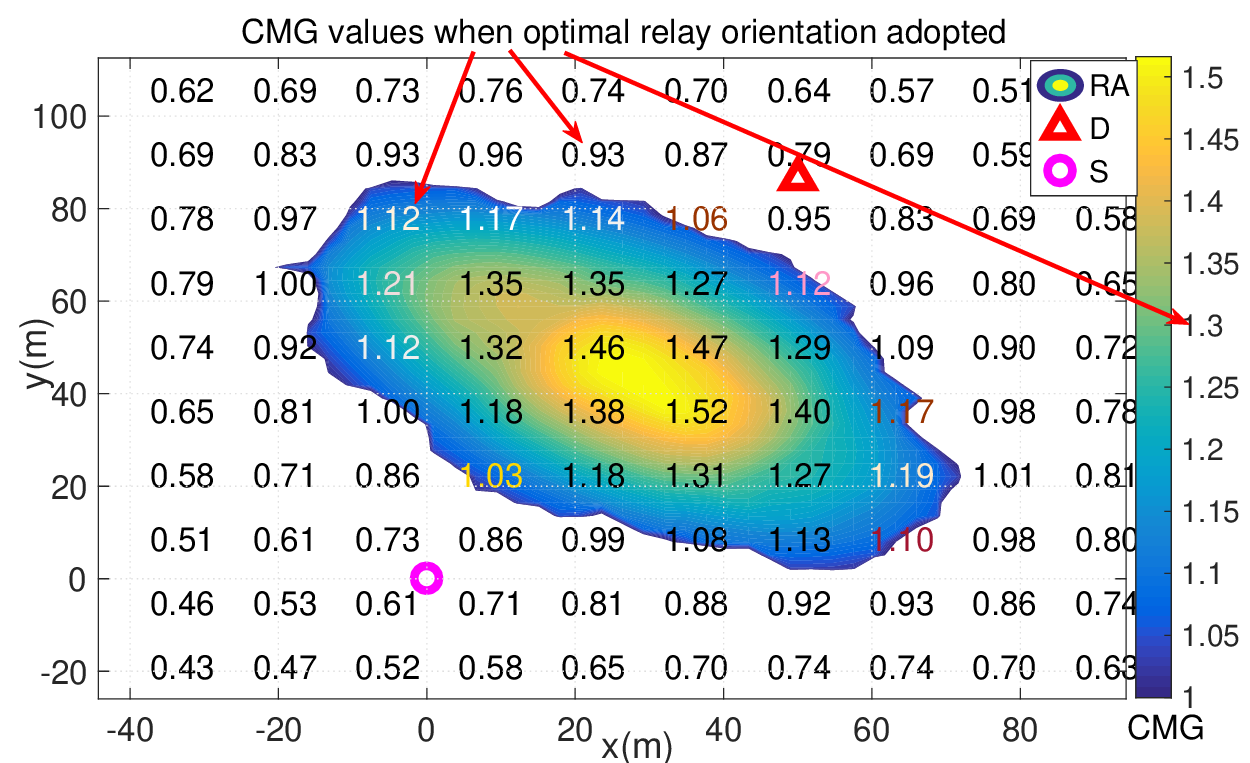}\\
        \caption{Effects of relay positions on the CMI performance. Here the optimal relay orientations for each position are roughly estimated by $\theta^*$ = $\arg\max\limits_{i\in[-90,89]}\mathcal{G}(x, y, i^\circ)$ where $i$ is an integer.
 }\label{fig_relayArea}
    \end{figure}

   \begin{figure}[t]
        \centering
        \includegraphics[width=3.2in]{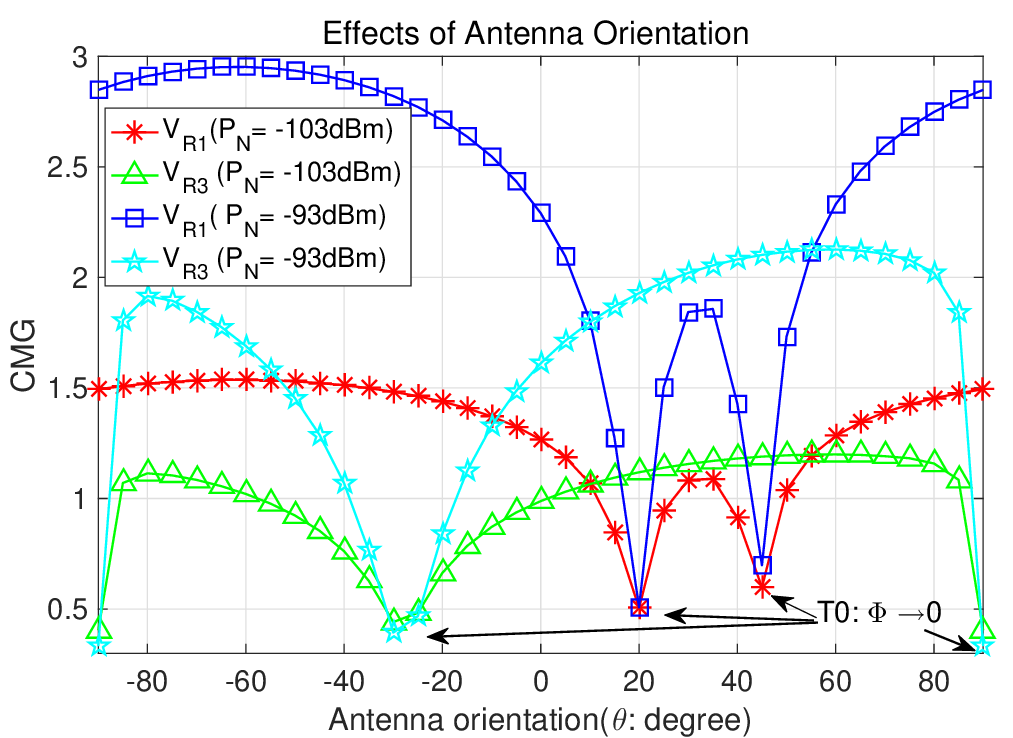}\\
        \caption{Effects of relay antenna orientation on the CMI performance for 2 relays with fixed positions ($\mathbf{v}_{R1}= (25,  43.3, \theta)$ and $\mathbf{v}_{R3}=(0, 86.6, \theta)$), respectively. $\Phi$ denotes the magnetic flux at receiver coils.
 }\label{fig_CMGTh}
    \end{figure}
In this section, we verify our analysis by simulating  the CMI channel bandwidth and RA. As depicted in\cite{Zhang2014Cooperative}, we assume the underground space is composed of dry soil whose conductivity  $\sigma= 0.01S\cdot$ m$^{-1}$, permeability is $\mu\!=\!4\pi$$\times 10^{-7}$ H$\cdot$m$^{-1}$ and permittivity value is $\epsilon\!= \!6.978 \cdot 10^{-11}$ F/m, hereby the skin depth $\delta\simeq 1/\sqrt{ \pi f \mu \sigma }$\cite{kisseleff2013channel}. The number of turns of antenna is $n\!=\!24$, antenna radius is $a\!=\!1$ m. A 17 AWG wire is used for the coil with the unit length resistance of $0.0166  \Omega/$m and $R_L$ is set to $R_0$\cite{lin2015distributed}. We set the resonance frequency $f_0\!=\!10$ KHz and transmitting power parameter $P_S\!=\!5$ W.  Our target distance between source and destination is 100 meters ($\mathbf{v}_d$ =$(50, 86.6, 30^\circ)$). The ambient noise level is assumed to $-103$ dBm\cite{Sun2010Magnetic}.

  In Fig.\ref{fig_bandwithP00}, the bandwidth $B_{AF}$ equals to $B1$ for the low transmitting power since $\mathcal{Z}_c^{-\frac{2}{3}}$ tends to 0.
  When the transmitting power is larger than 20 dBm, $B_{AF}$ tends to $B_1$ since $\mathcal{Z}_c^{-\frac{2}{3}}$ is close to a constant. Otherwise, $B_{AF}$ is smaller than $B_1$. For the most TTE scenarios, Fig.\ref{fig_bandwithP00} also implies  Remark 1 since the transmit power would be much larger than 20 dBm. Moreover, the AD has a significant effect on the CMI channel bandwidth.

 Fig.\ref{fig_relayArea} and Fig.\ref{fig_CMGTh} show the effects of ADs on CMI channel performance and roughly estimate the RA.
  Fig.\ref{fig_relayArea} shows that the approximately elliptical RA covers most of the intersection communication range of the source and destination where a relay yields an increase of 0--52\% in the data rate. The positions close to the long axis of RA achieve more data rate improvement than other positions,  which further implies Remark \ref{prop_optv} in the most TTE scenarios. Fig.\ref{fig_CMGTh} indicates that the antenna orientation is more sensitive to the transmission rate, especially near the T0 points since there is a magnetic field parallel to one of the receiver coils (R and D). Also, the data rate improvement is more significant in the system with weak signals than in the system with strong signals.

\section{Conclusion}\label{sect conclusions}
In this paper, we propose a CMI system with an AF MI relay to improve the data rate. We derive the expressions of mutual inductance \emph{w.r.t.} Cartesian coordinates and coil orientation. After that, we focus on the closed-form expression of CMI channel bandwidth and CMI achievable rate gain which can be used for relay selection. Simulation results exhibit the remarkable influence of the antenna position and orientation. With the adoption of a relay, the transmission rate increases dramatically in the low-rate MI network.

\ifCLASSOPTIONcaptionsoff
  \newpage
\fi



%
%

\bibliographystyle{IEEEtran} 
\bibliography{IEEEabrv,MIref}

%




\end{document}